\documentclass[%
 preprint,
 amsmath,amssymb,
 aps,
prd,
]{revtex4-1}

\usepackage{graphicx}
\usepackage{amssymb}
\usepackage{color}
\usepackage{epstopdf}
\usepackage{slashed}
\usepackage{amsmath,amsfonts,amsthm,bm}

\newcommand{\xp}{x^{+}}
\newcommand{\xm}{x^{-}}

\newcommand{\xbp}{\bar{x}^{+}}
\newcommand{\xbm}{\bar{x}^{-}}

\newcommand{\yp}{y^{+}}
\newcommand{\ym}{y^{-}}

\newcommand{\lav}{\left\langle}
\newcommand{\rav}{\right\rangle}

\newcommand{\xt}{\mathbf{x}}
\newcommand{\yt}{\mathbf{y}}

\newcommand{\pt}{\mathbf{p}}
\newcommand{\qt}{\mathbf{q}}

\newcommand{\kb}{\bar{k}}

\newcommand{\kt}{\mathbf{k}}
\newcommand{\kbt}{\bar{\mathbf{k}}}

\newcommand{\kp}{k^{+}}

\newcommand{\km}{k^{-}}

\newcommand{\Kt}{\mathbf{K}}

\newcommand{\bt}{\mathbf{b}}

\newcommand{\tr}{\text{tr}}

\usepackage{slashed}
\usepackage{xcolor}

\newcommand{\nn}{\nonumber}
\usepackage{amssymb}
\usepackage{amsmath}
\usepackage{epsfig}
\newcommand\beq{\begin{eqnarray}}
\newcommand\eeq{\end{eqnarray}}

\newcommand{\nslash}{\slashed{n}}
\newcommand{\kslash}{\slashed{k}}
\newcommand{\kbslash}{\slashed{\bar{k}}}
\newcommand{\Dslash}{\slashed{D}}

\newcommand{\xb}{\bar{x}}
\newcommand{\xbt}{\bar{\mathbf{x}}}

\newcommand{\Pt}{\mathbf{P}}
\newcommand{\Qt}{\mathbf{Q}}

\newcommand{\dkt}{\mathbf{\delta k}}

\newcommand{\rt}{\mathbf{r}}

\newcommand{\PsiBar}{\bar{\Psi}}

\begin{document}
\preprint{INT-PUB-18-055}

\title{Space-Time Picture of Baryon Stopping in the
Color-Glass Condensate}
\author{Larry D. McLerran}
\affiliation{%
Institute for Nuclear Theory, University of Washington, Seattle, WA 98195  
}%
\author{S\"{o}ren Schlichting}
\affiliation{
Fakult\"{a}t f\"{u}r Physik, Universit\"{a}t Bielefeld, D-33615 Bielefeld, Germany  
}%
\author{Srimoyee Sen}%
\affiliation{
Institute for Nuclear Theory, University of Washington, Seattle, WA 98195
}%

\date{\today}

\begin{abstract}We discuss baryon stopping in the Color Glass Condensate description of high energy scattering.
We consider the scattering of a distribution of valence quarks on an ultra-relativistic sheet of colored charge.  We compute the distribution 
of scattered quarks from a composite projectile, and calculate the baryon currents before and after the collisions and on an event by event basis. 
We obtain simple analytic estimates of the baryon number compression and rapidity shifts, which in the idealized case of plane wave scattering, produce results that agree with considerations of Anishetty-Koehler-McLerran\cite{Anishetty:1980zp}.
\end{abstract}


\maketitle

\section{Introduction}

The immediate goal of this paper is to formulate the problem of baryon number evolution in the context
of our modern understanding  of ultra-relativistic nuclear collisions.  Baryon number evolution was considered 
for such collisions in very early work concerning such collisions by Anishetty-Koehler-McLerrran\cite{Anishetty:1980zp}.
There it was shown that if the typical longitudinal rapidity boost given to  the baryons of the fragmentation region
is y, then the baryons become compressed by a factor of $e^y$.  In that paper, various estimates were given
for this factor, and it was argued that there was sufficient life time of the produced matter and energy density
to produce interesting new states of matter.

In this paper, we wish to generalize these considerations to the extreme high energy limit of ultra-relativistic 
collisions.  In this limit, we expect that typical momentum scales will be large compared to the QCD scale 
$\Lambda_{QCD}$, and that one should be able
to describe these collisions using the theory of the Color Glass Condensate\cite{McLerran:1993ni},\cite{McLerran:1993ka},
\cite{Iancu:2000hn},\cite{Iancu:2001ad}.  

Within the CGC formalism, several authors have studied the problem of quark production \cite{Martinez:2018tuf,Dumitru:2002qt,Dumitru:2005gt} in dilute-dense collisions, where a hard parton from the projectile is scattered of the dense color field of a large nucleus. Based on the usual momentum space description, the fragmentation region of these collisions, where the baryon number sits, has been analyzed in \cite{MehtarTani:2008qg,MehtarTani:2009dv,Duraes:2014jxa}, demonstrating interesting features such as geometric scaling. However, such calculations performed entirely in momentum space, implicitly assume that the matter produced in such collision does not undergo significant interactions in the final state and hence do not provide information on the space-time dynamics.

There are two major issues associated with the space-time description of the fragmentations region in the high energy limit .  The first,
which we shall address in this paper, is the space-time description of the baryon density and the color charge carried
by quarks (minus anti-quarks) in such collisions.  The next problem, that we shall address later, is the production of gluons
and their space-time evolution in the collisions.

There is also a broader issue to which the considerations of  this paper might ultimately be generalized \cite{Shen:2017ruz}.
At ultra-relativistic energies, there is a region of non-asymptotic energy where the center of mass energy is not high enough so that the 
baryon number separates between projectile and fragment.  For large enough nuclei, the intrinsic momentum scales associated with the 
Color Glass Condensate, $Q_{sat}$ should be large enough so that we can apply CGC methodology.  Nevertheless, the entanglement of 
final state interactions, combined with our as yet poor understanding of baryon number evolution and gluon production in the 
fragmentation region complicates this problem.  

In this paper we will consider scattering of a distribution of quarks on an ultra-relativistic nucleus, which is Lorentz contracted to an infinitesimal sheet of color charge along the $x^+$ direction $\xp = 0$ as illustrated in Fig. 1. Of course, the quarks will be required to have some Fermi momenta, which we shall ignore throughout this paper.  This is justified because
the saturation momentum will be taken to be very large compared to the typical Fermi momenta. We first consider the simplest case of a plane wave scattering, and estimate the baryon number compression and rapidity shift.


We then move on to a more involved modeling of the distribution of valence quarks in a composite projectile, to compute the space-time evolution of the baryon density in a realistic collision. 
We present our results for the current density as a function of the generalized parton distribution function and the beam function 
describing the distribution of the nucleons inside a nucleus. Finally we consider a realistic ansatz for these distribution functions 
and plot the resulting baryon current density. Our analysis shows that a classical particle picture adequately describes the baryon 
charge evolution after the scattering event.

\begin{figure}
\begin{center}

		\includegraphics[width=0.75\linewidth]{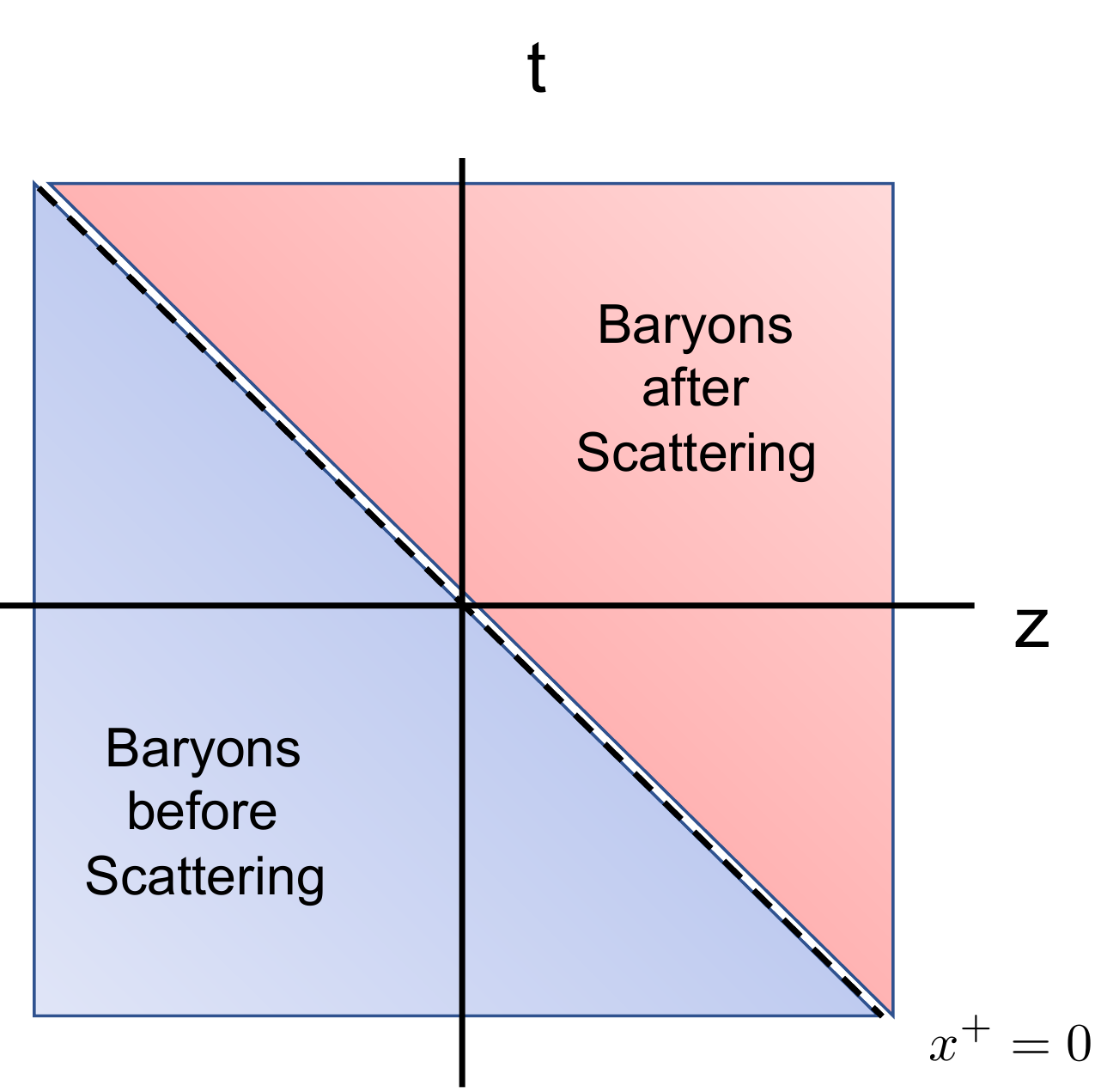}
\end{center}	

\caption{Scattering of a sheet infinite in the transverse direction on quarks initially at rest }
\label{sheetonB}
\end{figure}



This paper is organized as follows: We begin by formulating the baryon stopping problem in the color-glass condensate picture in Sec.~\ref{seq:Setup}. This is 
then followed by an analysis of a single quark plane-wave scattering off of a color charged sheet in Sec.~\ref{sqs}. We then analyze the scattering of a 
composite particle in Sec.~\ref{seq:CompProj}, finally specializing the analysis to describe the scattering of a large nucleus off of a sheet of color charge. We conclude with Sec.~\ref{sec:conc}.

\label{sec:Introduction}

\section{Setting up the  Calculation of Baryon Stopping}
\label{seq:Setup}
In this section we describe the various technical steps in calculating the resulting current density as a baryonic projectile crosses a sheet of colored glass corresponding to a nucleus. 
\subsection{Background gauge fields}
We begin with the classical Yang-Mills equations
\begin{eqnarray}
D_{\mu}F^{\mu\nu}=J^{\nu}+j^{\nu}\;,
\end{eqnarray}
in the presence of the strong source $J^{\nu}$ of the target nucleus and the weak source $j^{\nu}$ of the projectile nucleus where $D_{\mu}$ is the gauge covariant derivative. Specifying the gauge choice
\begin{eqnarray}
A^{+}=A_{-}=0\;,
\end{eqnarray}
the covariant current conservation law reads
\begin{eqnarray}
D_{\mu}J^{\mu}=0\;.
\end{eqnarray}
such that the color currents of target nucleus can then be described as
\begin{eqnarray}
J^{\nu}=\delta^{\nu-} ~g\rho(\xp,\xt)\;,
\end{eqnarray}
and will be treated non-perturbatively. Conversely the color currents $j^{\nu}$ associated with the projectile nucleus will be treated perturbatively, and do not contribute to the evolution of the baryon current at leading order.\\

Based on our gauge choice, the leading order solution to the gauge fields can be constructed as 
\begin{eqnarray}
-\nabla^2_\perp  A^{-} = g\rho(\xp,\xt)\;,
\end{eqnarray}
such that the gauge field can be chosen of the form
\begin{eqnarray}
A^{-}(\xp,\xm,\xt)= \frac{1}{-\nabla^2_\perp}~g\rho(\xp,\xt)\;,
\end{eqnarray}
and the field strength
\begin{eqnarray}
F^{i-}(\xp,\xm,\xt)=\frac{\partial^{i}}{-\nabla^2_\perp}~g\rho(\xp,\xt)\;,
\end{eqnarray}
is independent of $\xm$ and has support only in a small region of space-time concentrated around $\xp=0$. 

\subsection{Dirac equation at leading order} 
With the eventual goal of describing realistic collisions we now move on to analyze the propagation of fermions in the presence of the 
color field of the target nucleus. The fermion fields $\Psi$ satisfy the Dirac equation
\begin{eqnarray}
(i\Dslash-m)\Psi=0\;,
\end{eqnarray}
in the presence of the gauge fields $A^{\mu}$ sourced by the target nucleus. By performing the usual decomposition of the Dirac equation into the light-cone components
\begin{eqnarray}
\Psi^{+}=\frac{1}{2} \gamma^{+}\gamma^{-} \Psi\;, \qquad \Psi^{-}=\frac{1}{2} \gamma^{-}\gamma^{+} \Psi\;, 
\end{eqnarray}
the Dirac equation for the $\Psi^{+}$ component can be expressed as
\begin{eqnarray}
\Psi^{+}=\frac{1}{2i \partial_{-}} (i\gamma^{i}D_{i} +m) \gamma^{+} \Psi^{-}\;,
\end{eqnarray}
and becomes independent of the evolution time variable $\xp$. Based on this simplification, the relevant evolution equation in $\xp$ takes the form
\begin{eqnarray}
2i D_{+} \Psi^{-} = (i\gamma^{i}D_{i} +m) \frac{1}{i \partial_{-}}  (-i\gamma^{i}D_{i} +m) \Psi^{-}\;.
\end{eqnarray}

\subsection{Solution before the collision}
Outside the narrow region of support for the $A^{-}$ field, the equations of motion give rise to the on-shell condition
\begin{eqnarray}
(-2\partial_{+}\partial_{-} -\partial_{i}\partial^{i}-m^2)\Psi^{-}=-(\partial_{\mu}\partial^{\mu}+m^2) \Psi^{-}=0\;,
\end{eqnarray}
and the general solution in the region $\xp <0$ can be expressed as a super-position of plane wave modes
\begin{eqnarray}
&&\Psi(\xp<0,\xm,\xt)= \\
&& \quad (2\pi)^{-3} \sum_{s}\int_{0}^{\infty} \frac{dk^{+}}{2k^{+}} \int d^2\kt \left( u_{s}(k) b_{s,~in}(k) e^{-ikx} +  v_{s}(k) d^{\dagger}_{s,~in}(k) e^{+ikx} \right)\;, \nonumber
\end{eqnarray}
where $k^2=2k^{+}k^{-}-\kt^2=m^2$ is to be understood as an on-shell four momentum. We choose the Dirac spinors $u_{s}(k)$ and $v_{s}(k)$ to satisfy the following orthonormality
\begin{eqnarray}
 \bar{u}_{s}(k) \gamma^{+} u_{s'}(k)=2k^{+} \delta_{ss'}\;, \qquad  \bar{v}_{s}(k) \gamma^{+} v_{s'}(k)=2k^{+} \delta_{ss'}\;,
\end{eqnarray}
and completeness relations
\begin{eqnarray}
 \sum_{s} u_{s}(k) \bar{u}_{s}(k)=\kslash+m\;, \qquad  \sum_{s}v_{s}(k)\bar{v}_{s}(k)=\kslash-m\;,
\label{complete}
\end{eqnarray}
which we will use in the following to derive projection formulae and expressions for current matrix elements.

\subsection{Crossing the light-cone}
Crossing the light cone, the solution can be formally expressed as
\begin{eqnarray}
 && \Psi^-(\xp,\xm,\xt) \nonumber\\
&&= \mathcal{P} \exp \Big(i \int_{0^{-}}^{\xp} d\tilde{x}^{+} \Big[g A^{-}(\tilde{x}^{+},\xm\xt) 
- (i\gamma^{i}D_{i} +m) \frac{1}{2 i \partial_{-}}  (-i\gamma^{i}D_{i} +m)\Big] \Big)\nonumber \\ && \hspace{3.5 in}\Psi^-(\xp=0^{-},\xm,\xt)\;, \nonumber  \\ 
\end{eqnarray} 
which to leading order in the eikonal approximation yields the following crossing formula
\begin{eqnarray}
\Psi^-(\xp=0^{+},\xm,\xt) &=& V(\xt) \Psi^-(\xp=0^{-},\xm,\xt)\;,
\label{eik}
\end{eqnarray}
where $V(\xt)$ denotes the light-like Wilson line in the fundamental representation, i.e.
\begin{eqnarray}
V(\xt)=\mathcal{P} \exp \Big(ig \int_{0^{-}}^{0^{+}} d\xp~A^{-}(\xp,\xm\xt) \Big)\;.
\end{eqnarray}
\subsection{Solution after the collision}
Beyond $\xp>0^{+}$  the leading order solution is again of the plane wave form, i.e.
\begin{eqnarray}
\label{eq:solXP}
&&\Psi(\xp>0,\xm,\xt)=\\
&& \quad (2\pi)^{-3} \sum_{s}\int_{0}^{\infty} \frac{dk^{+}}{2k^{+}} \int d^2\kt \left( u_{s}(k) b_{s,out}(k) e^{-ikx} +  v_{s}(k) d^{\dagger}_{s,out}(k) e^{+ikx} \right)\;. \nonumber
\end{eqnarray}
However, the operators $b_{s,out}(k)$ and $d^{\dagger}_{s,out}(k)$ in the expansion of the fermion field now have to be obtained by matching to the solution on the $\xp >0$ side, i.e. right after the interacting with the color fields in the target nucleus to that for $\xp <0$, i.e. immediately before the interaction with the target nucleus. By applying the following reduction formula 
\begin{eqnarray}
\label{eq:bOut}
b_{s,out}(k)&=&\int d^2\yt \int d\ym~\bar{u}_{s}(k) \gamma^{+} \Psi(\yp=0^{+},\ym,\yt)~e^{+iky}\;, \\
\label{eq:dOut}
d^{\dagger}_{s,out}(k)&=&\int d^2\yt \int d\ym~\bar{v}_{s}(k) \gamma^{+} \Psi(\yp=0^{+},\ym,\yt)~e^{-iky}\;.
\end{eqnarray}
on the $\xp=0^{+}$ light-like hypersurface, it is straightforward to obtain the following momentum space expressions for $b_{s,out}(k)$ and $d^{\dagger}_{s,out}(k)$ 
\begin{eqnarray}
b_{s,out}(k)&=&\int \frac{d^2\pt}{(2\pi)^2}~\tilde{V}(\kt-\pt)~b_{s,in}(k^{+},\pt)\;,\\
d^{\dagger}_{s,out}(k)&=&\int \frac{d^2\pt}{(2\pi)^2}~\tilde{V}(\pt-\kt)~d^{\dagger}_{s,in}(k^{+},\pt)\;. 
\end{eqnarray}
which are frequently used in calculations of partonic cross-sections in the color glass condensate framework.  However, for our purpose of calculating the space-time evolution of baryon currents, it is in fact more convenient to work directly with the formal expressions in Eqns.~(\ref{eq:bOut},\ref{eq:dOut}). By inserting Eqns.~(\ref{eq:bOut},\ref{eq:dOut}) into Eq.~(\ref{eq:solXP}) and making use of the relation in Eq.~(\ref{complete}), we then obtain the general solution in the forward light-cone as
\begin{eqnarray}
\label{eq:solXPexplicit}
&& \Psi(\xp>0,\xm,\xt)= (2\pi)^{-3} \int_{0}^{\infty} \frac{dk^{+}}{2k^{+}} \int d^2\kt \int d\ym \int d^2\yt \\
&& \quad \left[ (\kslash+ m) e^{-i k(x-y)} + (\kslash- m) e^{+i k(x-y)}\right]_{y^{+}=0} ~V(\yt)\gamma^{+}~\Psi(\yp=0^{-},\ym,\yt)\;.\nonumber
\end{eqnarray}
We note that the right hand side expression only involves the fermion field $\Psi(\yp=0^{-},\ym,\yt)$ immediately before the interaction with the target nucleus. Conversely, the subsequent propagation in the color field of the target and after the collision are expressed via the convolution in coordinate space with the forward scattering amplitude $V(\yt)\gamma^{+}$ and the free-fermion propagator
\begin{eqnarray}
&&G(\xp>0,\xm,\xt|\yp=0^{+},\ym,\yt)=  \\
&& \qquad (2\pi)^{-3} \int_{0}^{\infty} \frac{dk^{+}}{2k^{+}} \int d^2\kt \left[ (\kslash+ m) e^{-i k(x-y)} + (\kslash- m) e^{+i k(x-y)}\right]_{y^{+}=0}\;. \nonumber
\end{eqnarray}
\subsection{Calculation of baryon current}
Equipped with the explicit leading order solution of the fermion field in the forward light cone in Eq.~(\ref{eq:solXPexplicit}), we can now proceed to calculate the vector current 
\begin{eqnarray}
j^{\mu}(y)=\left\langle  \PsiBar(y) \gamma^{\mu} \Psi(y) \right\rangle\;.
\end{eqnarray}
Based on the above expressions the vector current in the forward light-cone can be compactly expressed as
\begin{eqnarray}
&& \lav j^{\mu}(y) \rav = \nonumber\\
&&\frac{1}{(2\pi)^6}\int_{-\infty}^{\infty} \frac{dk^{+}}{2k^{+}} \int_{-\infty}^{\infty} \frac{d\kb^{+}}{2\kb^{+}}  \int d^2\kt \int d^2\kbt \int d\xm \int d\xbm \int d^2\xt \int d^2\xbt \nonumber \\
&&\lav \PsiBar(\xbp=0^{-},\xbm,\xbt)~V^{\dagger}(\xbt)~
 \Gamma^{\mu}(k,\kb) ~V(\xt)~\Psi (\xp=0^{-},\xm,\xt) \rav e^{-i \kb(\xb-y)+i k(x-y)} \; \nonumber 
\end{eqnarray}
where we denote the current matrix element as
\begin{eqnarray}
\Gamma^{\mu}(k,\kb)=\nslash(\kbslash+m) \gamma^{\mu}  (\kslash+m)  \nslash\;,
\end{eqnarray}
Note that  in the above expression, the integrals over $k^{+},\kb^{+}$ now extend from $-\infty$ to $+\infty$ with the negative region associated with the anti-particle contribution; $k^2=\kb^2=m^2$ are to be understood as on-shell four momenta and $n^{\mu}$ denotes the light like four vector  $n^{\mu}=(n^{+},n^{-},\mathbf{n})=(0,1,0,0)$ such that $\nslash=\gamma^{+}\;, n^2=n_{\mu}n^{\mu}=0$ and  $n_{\mu}k^{\mu} = k^{+}$. Decomposing the expectation value into color singlet structures in the projectile and target nucleus by using 
\begin{eqnarray}
\lav \Big(V^{\dagger}(\xbt) V(\xt)\Big)_{ij} \rav= \delta_{ij} D(\xt,\xbt)\;, \qquad  D(\xt,\xbt)=\frac{1}{N_c} \tr [ V^{\dagger}(\xbt) V(\xt)]
\end{eqnarray}
the expression for the vector current assumes the following factorized form
\begin{eqnarray}
&&\lav j^{\mu}(y) \rav \nn\\
&&=(2\pi)^{-6} \int_{-\infty}^{\infty} \frac{dk^{+}}{2k^{+}} \int_{-\infty}^{\infty} \frac{d\kb^{+}}{2\kb^{+}}  \int d^2\kt \int d^2\kbt \int d\xm \int d\xbm \int d^2\xt \int d^2\xbt \nonumber \\
&& D(\xt,\xbt)~\lav \PsiBar(\xbp=0^{-},\xbm,\xbt)~
 \Gamma^{\mu}(k,\kb) ~\Psi (\xp=0^{-},\xm,\xt) \rav e^{-i\kb(\xb-y)} e^{+i k(x-y)}\;.  \nonumber \\ 
\end{eqnarray}
Evaluating the current matrix element explicitly according to
\begin{eqnarray}
\Gamma^{\mu}(k,\kb) &=& \nslash \left[ 2k^{\mu} (\kb \cdot n) +2 \kb^{\mu} (k \cdot n) - 2 n^{\mu} (k\cdot \kb -  m^2) \right] \nn\\
&+& 2i \gamma^{\nu}\gamma^{5} \left[\epsilon^{\mu\nu\alpha\beta} \Big( (\kb \cdot n) k_{\alpha} n_{\beta} - (k \cdot n) \kb_{\alpha} n_{\beta} \Big) - n^{\mu} \epsilon^{\nu\alpha\beta\gamma} k_{\alpha} \kb_{\beta} n_{\gamma}  \right]  \nonumber \\
&+& 2im~\sigma^{\mu\alpha} n_{\alpha} \Big((k \cdot n) - (\kb \cdot n) \Big) + 2im~\sigma^{\alpha\beta} n_{\beta} \Big(  \kb_{\alpha} -  k_{\alpha}\Big) \nonumber 
\end{eqnarray}
one recognizes three distinct contributions related to the different operator structures of the hadronic matrix element.  Clearly, the presence of the hadronic correlation function in conjunction with the oscillating phase factors associated with the propagation of quarks after the collision can give rise to a complicated space-time structure of the currents. 

\section{Baryon stopping for a single quark}\label{sqs}
In order to obtain an intuitive insight into the dynamics, we will therefore first discuss the case of a single quark scattering of the color fields of a dense nucleus.  One finds that for the scattering of a single quark in a momentum eigenstate only the unpolarized contribution of the current matrix element survives. Evaluating the corresponding correlation function of the quark fields for $\xp\leq 0^{-}$ according to
\begin{eqnarray}
\lav \PsiBar(\xb) \nslash \Psi(x) \rav = \int \frac{d^3\bar{p}}{(2\pi)^3} \int \frac{d^3p}{(2\pi)^3} \sum_{s\bar{s}} \frac{ \bar{u}_{\bar{s}}(\bar{p}) \nslash u_{s}(p)}{\sqrt{2E_{\bar{p}}} \sqrt{2E_{p}}}~\lav b^{\dagger}_{\bar{s}}(\bar{p}) b_{s}(p) \rav~ e^{-ipx}e^{+i\bar{p}\xb}, \nonumber \\
\end{eqnarray}
for a momentum eigenstate with on-shell four momentum $P=(P^{+},P^{-},\Pt)$ and averaging over spin and color according to
\begin{eqnarray}
\lav b^{\dagger}_{\bar{s}}(\bar{p}) b_{s}(p) \rav = (2\pi)^{3} \delta^{(3)}(p-P)  (2\pi)^{3} \delta^{(3)}(\bar{p}-P) \frac{\delta_{s\bar{s}}}{2} ~\mathcal{N} 
\end{eqnarray}
where $\mathcal{N}$ is a constant for the normalization of the plane wave states, we get
\begin{eqnarray}
\lav \PsiBar(\xbp=0^{-},\xbm,\xbt) \nslash \Psi(\xp=0^{-},\xm,\xt) \rav = \mathcal{N} \frac{(P\cdot n)}{E_P}~e^{-iPx} e^{+iP\xb}\;.
\end{eqnarray}
By integrating the expression for the current over $x^{-}$ and $\xb^{-}$ one readily obtains 
\begin{eqnarray}
\lav j^{\mu}(y) \rav &=& \frac{\mathcal{N}}{(2\pi)^4} \int d^2\kt \int d^2\kbt \frac{1}{2E_P} \left(\kb^{\mu} + k^{\mu} - \frac{k\cdot \kb -m_q^2}{P^{+}} n^{\mu}\right)_{k^{+}=\kb^{+}=P^{+}} \nonumber \\ 
&& \int d^2\xt \int d^2\xbt~D(\xt,\xbt)~e^{-i \kt(\xt-\yt)} e^{+i\kbt(\xbt-\yt)}~e^{+i\Pt(\xt-\xbt)}~e^{-i(k^{-}-\kb^{-})y^{+}} \;,\nonumber \\
\end{eqnarray}
By evaluating the contributions to the different components of the current explicitliy according to
\begin{eqnarray}
\kb^{-} + k^{-} - \frac{k\cdot \kb -m^2}{P^{+}}=\frac{\kt \kbt +m^2}{2P^{+}} = \frac{ \Big(\frac{\kt+\kbt}{2}\Big)^2 + m^2}{2P^{+}} - \frac{(\kt-\kbt)^2}{4}\;, 
\end{eqnarray}
and simplifying the phase factor in the exponential using
\begin{eqnarray}
\kb^{-} - k^{-}=\frac{\kt^2 -\kbt^2}{2P^{+}} = \frac{ \Big(\frac{\kt+\kbt}{2}\Big)}{P^{+}}(\kt-\kbt) \;, 
\end{eqnarray}
most of the integrals can be performed by using a change of variables to sum and difference coordinates, and it is then straightforward to obtain the final result
\begin{eqnarray}
\label{eq:SingleQuarkCurrents}
\lav j^{\mu}(y) \rav &=&   \mathcal{N}   \int \frac{d^2\Kt}{(2\pi)^2}~\frac{1}{E_P} \begin{pmatrix} P^{+} \\  \Kt^{i} \\ \frac{ \Kt^2+m_q^2}{2P^{+}} + \frac{\nabla_{\yt}^2}{4} \end{pmatrix}~D_{W}\Big(\Kt-\Pt,\yt -\frac{\Kt}{P^{+}} y^{+}\Big)\;, \nonumber \\
\end{eqnarray}
where $D_{W}(\Qt,\bt)$ denotes the Wigner transform of the dipole scattering amplitude
\begin{eqnarray}
\label{eq:DipWig}
D_{W}(\Qt,\bt)=\int d^2\rt~D(\bt+\rt/2,\bt-\rt/2)~e^{-i\Qt\rt} \;.
\label{dip}
\end{eqnarray}
Note that the above expression for the baryon current has a straightforward interpretation in terms of classical particle picture: Incoming quarks with four momentum $P=(P^{+},\frac{\Pt^2+m_q^2}{2P^{+}},\Pt)$ interact with the nucleus at a space-time position $x^{\mu}=(0,\xm,\xt)$ and receive a momentum transfer given by $(0,\frac{(\Kt-\Pt)^2-\Pt^2}{2P^{+}},\Kt-\Pt)$. Subsequently for $\yp>0$ they propagate freely from the space time point $x^{\mu}=(0,\xm,\xt)$ of the interaction to the space time point $y^{\mu}=(\yp,\ym,\yt=\xt +\frac{\Kt}{P^{+}} y^{+})$ where the current is measured.Since we considered the scattering of momentum eigenstates, the interaction point is delocalized in $\xm$ and the resulting baryon current in Eq.~(\ref{eq:SingleQuarkCurrents}) becomes independent of $\ym$. 

Notably the formula in Eq.~(\ref{eq:SingleQuarkCurrents}) is extremely useful to estimate the effects of baryon number compression and baryon stopping. Neglecting the impact parameter ($\bt$) dependence in the target, and considering the simplest possible ansatz for the Dipole scattering amplitude
\begin{eqnarray}
D(\Kt,\bt)=\frac{(2\pi)^2}{\pi Q_s^2} \exp\left(-\frac{\Kt^2}{Q_s^2}\right)
\end{eqnarray}
where $Q_{s}$ is the saturation momentum of the nucleus, one readily obtains the result
\begin{eqnarray}
\lav j^{+}\rav_{\yp>0} &=& \lav j^{+}\rav_{\yp<0}\;,  \\
 \lav j^{-}\rav_{\yp>0} &=& \frac{Q_s^2}{m_T^2} \lav j^{-}\rav_{\yp<0}\;,
\end{eqnarray}
where $\lav j^{\pm}\rav_{\yp<0}$ and $\lav j^{\pm}\rav_{\yp>0}$ describe the currents before and after the interaction with the nucleus, and $m_T^2=\Pt^2+m_q^2$ denotes the transverse mass of the projectile. Decomposing the current $j^{\mu}$ in terms of the rest-frame density $n$ and flow velocity $u^{\mu}=\frac{1}{\sqrt{2}} (e^{+y},e^{-y},{\bf 0})$ according to\footnote{Note that strictly speaking, for a single quark with definite momentum $\Pt$ the transverse components of the current $j^{i}$ are also non-vanishing. However, generalizing the above expressions to include an average over $\Pt\sim \Lambda_{QCD}$, one finds that the transverse components $j^{i}$ vanish due to rotational symmetry and we will therefore neglect them in our analysis.} 
\begin{eqnarray}
j^{\mu}=n u^{\mu}
\end{eqnarray}
one finds that the rapidity is shifted by
\begin{eqnarray}
  \lav y \rav_{\yp>0} -  \lav {y} \rav_{\yp<0} =  \frac{1}{2} \log \frac{Q_s^2}{m_T^2}\;,
\end{eqnarray}
while the  density gets compressed by a factor $\frac{Q_s}{m_T}$, i.e.
\begin{eqnarray}
\lav n \rav_{\yp>0} = \frac{Q_s}{m_T} \lav n \rav_{\yp<0}\;.
\end{eqnarray}
This is the result of Anishetty-Koehler-Mclerran (AKM), except that in our analysis compression factor is determined by the dynamics of the Color Glass Condensate. We have an expression for the current density valid configuration by configuration. There are several important distinctions from the AKM treatment:  The CGC allows one to handle spatial fluctuations. Most importantly, the typical transverse momentum scale is not the QCD scale, as was assumed in AKM, but is the saturation momentum of the projectile nucleus which can be much larger than the QCD scale and grows with increasing beam energy.

\section{Baryon stopping for composite projectiles}
\label{seq:CompProj}
We will now consider the more realistic scenario of a composite projectile scattering of the color field of a the target nucleus. We focus for simplicity on the unpolarized contribution given by \footnote{Note that in Eq.~(\ref{crnt}) and all subsequent expressions, we denote $x=(0,\xm,\xt)$ and $\xb=(0,\xbm,\xbt)$, such that the phase factor in the exponentials are to be evaluated at $\xp=\xbp=0$.}
\begin{eqnarray}
&& \lav j^{\mu}(y) \rav 
=(2\pi)^{-6} \int_{-\infty}^{\infty}  \frac{dk^{+}}{2k^{+}} \int_{-\infty}^{\infty} \frac{d\kb^{+}}{2\kb^{+}}  \int d^2\kt \int d^2\kbt \int d^2\xt \int d^2\xbt \nonumber \\
&& \qquad D(\xt,\xbt)~\left[ 2k^{\mu} (\kb \cdot n) +2 \kb^{\mu} (k \cdot n) - 2 n^{\mu} (k\cdot \kb -  m^2) \right]  \nonumber \\
&& \int d\xm \int d\xbm \lav \PsiBar(\xbp=0^{-},\xbm,\xbt)~\nslash ~\Psi (\xp=0^{-},\xm,\xt) \rav e^{-i \kb(\xb-y)} e^{+ik(x-y)}\;.  \nonumber \\ 
\label{crnt}
\end{eqnarray}
We denote the relevant quark distribution in the projectile as
\begin{eqnarray}
&& q(k^{+},\kb^{+},\qt,\bt)= \int d^2\rt~\int d\xm \int d\xbm ~e^{-i\kb^{+} \xb^{-}} e^{+i k^{+}x^{-}} e^{- i \qt\rt} \\
&& \qquad \qquad  \lav \PsiBar(\xb^{+}=0,\xbm,\bt-\rt/2)~ \nslash ~\Psi (\xp=0,\xm,\bt+\rt/2) \rav \nonumber
\label{qdist}
\end{eqnarray}
and define
\begin{eqnarray}
q(k^{+},\kb^{+},\bt)&\equiv&\int \frac{d^2\qt}{(2\pi)^2}q(k^{+},\kb^{+},\qt,\bt)\;,\nonumber\\
&=& \int d\xm \int d\xbm ~e^{-i\kb^{+} \xb^{-}} e^{+i k^{+}x^{-}}\nn\\
&&  \lav \PsiBar(\xb^{+}=0,\xbm,\bt)~ \nslash ~\Psi (\xp=0,\xm,\bt) \rav \nonumber\;.
\end{eqnarray}

Now in order to localize the interaction point in both $\xm$ and $\xt$, we consider a wave-package of hadrons with momentum distribution $\Omega(p,\bar{p})$ such that the operator expectation value is given by
\begin{eqnarray}
&& q(k^{+},\kb^{+},\bt)=    \nn \\
&& \int \frac{d^4p}{{(2\pi)^4}} \int \frac{d^4\bar{p}}{{(2\pi)^4}} ~ (2\pi)^2 \delta(p^2-M_N^2)   \delta(\bar{p}^2-M_N^2)~\Omega(p,\bar{p})  \int d X^{-}  e^{i(\kb^{+}-k^{+})X^{-}}   \nonumber \\
&& \int d\delta\xm e^{i\frac{\kb^{+}+k^{+}}{2}\delta\xm}\lav \bar{p} | \PsiBar(0,X^{-}-\delta\xm/2,\bt)~ \nslash ~\Psi (0,X^{-}+\delta\xm/2,\bt) | p \rav \nonumber\;,
\end{eqnarray}
where $\Omega(p,\bar{p})$ is normalized such that
\begin{eqnarray}
\int \frac{d^4p}{{(2\pi)^4}} \int \frac{d^4\bar{p}}{{(2\pi)^4}} ~ (2\pi)^2 \delta(p^2-M_N^2)   \delta(\bar{p}^2-M_N^2)~\Omega(p,\bar{p}) \lav \bar{p} || p \rav =1
\end{eqnarray}
By performing a shift of the arguments of the hadronic matrix elements according to
\begin{eqnarray}
&&\lav \bar{p} | \PsiBar(0,X^{-}-\delta\xm/2,\bt)~ \nslash ~\Psi (0,X^{-}+\delta\xm/2,\bt) | p \rav =  \nonumber  \\
 && \quad e^{-i(\bar{p}^{+}-p^{+})X^{-}} e^{i(\bar{\pt}-\pt)\bt}  \lav \bar{p} \left| \PsiBar\Big(0,-\frac{\delta\xm}{2},0\Big)~ \nslash ~\Psi \Big(0,+\frac{\delta\xm}{2},0\Big) \right| p \rav 
\end{eqnarray}
one recognizes that the hadronic matrix element can be expressed in terms of a generalized parton distribution (GPD) (see e.g. \cite{Diehl:2003ny} for a comprehensive review)
\begin{eqnarray}
q^{V}(x,\xi,t)=\frac{1}{2} \int \frac{d\delta\xm}{(2\pi)} e^{i\frac{\kb^{+}+k^{+}}{2}\delta\xm} \lav \bar{p} \left| \PsiBar\Big(0,-\frac{\delta\xm}{2},0\Big)~ \nslash ~\Psi \Big(0,+\frac{\delta\xm}{2},0\Big) \right| p \rav\;, \nonumber  
\end{eqnarray}
which depends on the variables
\begin{eqnarray}
x=\frac{\kb^{+}+k^{+}}{p^{+}+\bar{p}^{+}} \qquad \xi=\frac{p^{+}-\bar{p}^{+}}{p^{+}+\bar{p}^{+}} \qquad t=(\bar{p}-p)^2
\end{eqnarray}
By performing also the integration over $X^{-}$, which sets the longitudinal momentum difference of the hadron $\bar{p}^{+}-p^{+}$ equal to that of the parton $\kb^{+}-k^{+}$, the relevant distribution is then given by
\begin{eqnarray}
q(k^{+},\kb^{+},\bt)&=&   4\pi~\int \frac{d^4p}{{(2\pi)^4}} \int \frac{d^4\bar{p}}{{(2\pi)^4}} ~ (2\pi)^2 \delta(p^2-M_N^2) \delta(\bar{p}^2-M_N^2)~\Omega(p,\bar{p}) \nonumber \\
&&\qquad~e^{i(\bar{\pt}-\pt)\bt}~q^{V}(x,\xi,t)~(2\pi)\delta\Big((\kb^{+}-k^{+}) -(\bar{p}^{+}-p^{+})\Big) \;. \nonumber \\
\end{eqnarray}
Note that physically, the distribution $q^{V}(x,\xi,t)$ can be viewed as the probability to emit a quark with momentum fractions
\begin{eqnarray}
x+\xi=2k^{+}/(p^{+}+\bar{p}^{+})\;, \qquad x-\xi=2\kb^{+}/(p^{+}+\bar{p}^{+})\;.
\end{eqnarray}
in the amplitude and in the complex conjugate amplitude respectively, such that in contrast to the single quark scattering example, the longitudinal momenta $k^{+}$ and $\kb^{+}$ are in general different. We will see shortly, how this will lead to modifications of the baryon currents in the forward light-cone.\\

Equipped with the relevant hadronic matrix element, we can now turn back to the evaluation of the baryon current. Defining center of mass and difference coordinates for the transverse variables according to
\begin{eqnarray}
\bt=\frac{\xt+\xbt}{2}\;, \quad \rt=\xt-\xbt\;, \quad \Kt=\frac{\kt +\kbt}{2}\;, \quad \dkt=\kt-\kbt\;,
\label{avg}
\end{eqnarray}
the phase factors become
\begin{eqnarray}
e^{-i \kb(\xb-y)} e^{+i k(x-y)} &=& e^{-i \kb^{+} (\xb^{-}-y^{-})} e^{+i  k^{+}(x^{-}-y^{-})} e^{-i\Kt \rt -\dkt (\bt-\yt)} \\
&& e^{i \Big(\Kt^2+\frac{\dkt^2}{4}+m^2\Big) \Big(\frac{1}{2 \kb^{+}} - \frac{1}{2 k^{+}}\Big)y^{+}} e^{-i \Kt \dkt  \Big(\frac{1}{2 \kb^{+}} + \frac{1}{2k^{+}}\Big) y^{+}} \nonumber.
\end{eqnarray}
and the different components of the current matrix element can be evaluated according to
\begin{eqnarray}
&&2k^{+} (\kb \cdot n) +2 \kb^{+} (k \cdot n)= 4k^{+}\kb^{+}\;, \nonumber
\end{eqnarray}
\begin{eqnarray}
&& 2k^{i} (\kb \cdot n) +2 \kb^{i} (k \cdot n)= 4k^{+}\kb^{+} \left( \Kt^{i} \Big(\frac{1}{2\kb^{+}} + \frac{1}{2 k^{+}}\Big) - \dkt/2\Big(\frac{1}{2\kb^{+}} - \frac{1}{2 k^{+}}\Big) \right) \;, \nonumber
\end{eqnarray}
\begin{eqnarray}
&&2k^{-} (\kb \cdot n) +2 \kb^{-} (k \cdot n) - 2 n^{-} (k\cdot \kb - m^2) = 2 \left( \Kt^2 - \frac{\dkt^2}{4} +m^2 \right)\;. \nonumber
\end{eqnarray}
By switching to the center of mass and difference coordinates as defined in Eq.~(\ref{avg}) and performing some of the integrations, we then obtain
\begin{eqnarray}
&& \begin{pmatrix} \lav j^{+}(y) \rav  \\ \lav j^{i}(y) \rav \\ \lav j^{-}(y) \rav \end{pmatrix}
=(2\pi)^{-8} \int_{-\infty}^{\infty} dk^{+} \int_{-\infty}^{\infty} d\kb^{+} \int d^2\Kt \int d^2\dkt \int d^2\bt \int d^2\qt \nonumber \\
&& \quad~D_{W}(\Kt-\qt,\bt)~q(k^{+},\kb^{+},\qt,\bt)  
\begin{pmatrix} 1 \\
 \Kt^{i} \Big(\frac{1}{2\kb^{+}} + \frac{1}{2 k^{+}}\Big) - \dkt/2\Big(\frac{1}{2\kb^{+}} - \frac{1}{2 k^{+}}\Big) \\
\frac{ \Kt^2 - \frac{\dkt^2}{4} +m^2 }{2k^{+}\kb^{+}}
\end{pmatrix}
\nonumber \\
&&  e^{+i  (\kb^{+}-k^{+}) y^{-}} e^{-i\dkt (\bt-\yt)} e^{+i \Big(\Kt^2+\frac{\dkt^2}{4}+m^2\Big) \Big(\frac{1}{2 \kb^{+}} - \frac{1}{2 k^{+}}\Big)y^{+}} e^{-i \Kt \dkt  \Big(\frac{1}{2 \kb^{+}} + \frac{1}{2k^{+}}\Big) y^{+}}\;, \nonumber \\
\end{eqnarray}
where $D_{W}$ is the Wigner transform of the dipole distribution of the target defined in Eq.~(\ref{dip}) and $q(k^{+},\kb^{+},\qt,\bt)$ denotes the hadronic correlation function of the projectile defined in Eq.~(\ref{qdist}). Since the  intrinsic transverse momentum of the quarks in the projectile $\qt \sim \Lambda_{QCD} $, whereas the typical momentum transfer $\Kt \sim Q_s \gg \Lambda_{QCD} $, we can further neglect the intrinsic transverse momentum of the quark to obtain
\begin{eqnarray}
&& \begin{pmatrix} \lav j^{+}(y) \rav  \\ \lav  j^{i}(y) \rav  \\ \lav j^{-}(y) \rav  \end{pmatrix}
=\int \frac{d^4p}{{(2\pi)^4}} \int \frac{d^4\bar{p}}{{(2\pi)^4}} ~ (2\pi)^2 \delta(p^2-M_N^2)   \delta(\bar{p}^2-M_N^2)~\Omega(p,\bar{p})  \nonumber \\
&&2P^{+}~\int_{-\infty}^{\infty} dx~q^{V}(x,\xi,t)~\int \frac{d^2\Kt}{(2\pi)^2} \int \frac{d^2\dkt}{(2\pi)^2} \int d^2\bt~D_{W}(\Kt,\bt)~\nn\\
&&\,\,\,\,\,\,\,\,\,\,\,\,\,\,\,\,\,\,\,\,\,\,\,\,\,\,\,\,\,\,\,\,\,\,\,\,\,\,\,\frac{1}{xP^{+}}
\begin{pmatrix} 
xP^{+} \\
\left( \Kt^{i} + \left( \frac{-\xi}{x} \right) \frac{\dkt^{i}}{2} \right) \frac{x^2}{x^2-\xi^2}\\
\frac{\Kt^2 - \frac{\dkt^2}{4} +m^2}{2xP^{+}}  \frac{x^2}{x^2-\xi^2}
\end{pmatrix} 
\nonumber \\
&&  e^{i(\bar{\pt}-\pt)\bt} e^{+2i  \left( \frac{-\xi}{x} \right) \left(x P^{+} y^{-} - \frac{\Kt^2+\frac{\dkt^2}{4}+m^2}{2xP^+}  \frac{x^2}{x^2-\xi^2} y^{+} \right)} e^{+i\dkt \left(\yt-\bt - \frac{\Kt}{xP^{+}}   \frac{x^2}{x^2-\xi^2} y^{+} \right)}\;, \nonumber \\
\label{curr1}
\end{eqnarray}
where $P^{+}=(p^{+}+\bar{p}^{+})/2$, $\xi=(p^{+}-\bar{p}^{+})/(p^{+}+\bar{p}^{+})$ and $t=(\bar{p}-p)^2$. Even though the expressions for the currents simplify dramatically in the limit $\yp=0$ i.e. immediately after the collision, where e.g. $j^{+}$ is simply related to the Dirac and Pauli form factors (see \ref{A})
\begin{eqnarray}
 \left. \lav j^{+}(y) \rav \right|_{y^{+}=0} &=& \int \frac{d^4p}{{(2\pi)^4}} \int \frac{d^4\bar{p}}{{(2\pi)^4}} ~ (2\pi)^2 \delta(p^2-M_N^2)   \delta(\bar{p}^2-M_N^2)~\Omega(p,\bar{p})  \nonumber \\
&&\bar{u}(\bar{p})\left[ F_{1}^{q}(t) \gamma^{+} +F_{2}^{q}(t) \frac{i \sigma^{+\alpha} (\bar{p}-p)_{\alpha}}{2M_{N}}  \right] u(p)~e^{i\delta p y} 
\end{eqnarray}
we find that the most general result in Eq.~(\ref{curr1}) no longer admits a simple probabilistic interpretation in terms of classical particles. Because the longitudinal momenta of the quark/anti-quark are in general different in the amplitude and conjugate amplitude, interference effects can play an important role in determining the space-time structure of currents in the forward light-cone. Noteably, these effects can be understood as a consequence of the uncertainty principle which prohibits a precise simultaneous determination of the interaction point $\xm$ and the momentum $\kp$. However, as we will see shortly, such effects are suppressed in the collision of large nuclei, where a semi-classical picture can be recovered by exploiting the separation of scales between the size of the nucleus and the size of the nucleon.  \\

\subsection{Baryon stopping in collision of large nuclei}
We now focus specifically on the case of baryon stopping in the collision of large nuclei, for which some important simplifications of the general result occur. We consider the projectile nucleus as composed of individual nucleons and adopt the semi-classical treatment of Kovchegov and Sievert\cite{Kovchegov:2015zha}  for the position and momentum distribution $\Omega(p,\bar{p})$ of nucleons inside the projectile
\begin{eqnarray}
\Omega(p,\bar{p})=\sqrt{4 p^{+}\bar{p}^{+}}~\rho_{N}(t)~(2\pi)^2\delta^{(2)}\left( \Pt \right)~(2\pi)\delta\left(P^{+}-P^{+}_{N}\right)\;.
\label{omega}
\end{eqnarray}
We note that in Eq.~(\ref{omega}), we neglected the long. and transverse momentum uncertainty of the nucleons, anticipating that the motion of nucleons inside the nucleus is non-relativistic such that $|\Pt| \ll P^{+}_{N}$ and similarly $|P^{+} - P^{+}_{N}|  \ll  P^{+}_{N}$. By $\rho_{N}(t)$ we denote the Fourier transform of the density of nucleons inside the projectile, which is constrained by rotational invariance in the Breit frame (B.F.) where
\begin{eqnarray}
\tilde{\rho}_{N}(t)\stackrel{B.F.}{=} \int d^3b~\rho_{N}(b)~e^{-iqb}\;, \qquad t\stackrel{B.F.}{=}-q^2\;.
\end{eqnarray}
By using the explicit expression for $t$ in our usual frame
\begin{eqnarray}
t=-\frac{4M_N^2 \xi^2}{1-\xi^2} - \frac{(1+\xi) \pt - (1-\xi) \bar{\pt}}{1-\xi^2}\stackrel{\Pt=0}{=} -\frac{4M_N^2 \xi^2}{1-\xi^2} - \frac{\delta\pt^2}{1-\xi^2}
\end{eqnarray}
one can then express the averages over the projectile nucleon states as
\begin{eqnarray}
&& \int \frac{d^4p}{{(2\pi)^4}} \int \frac{d^4\bar{p}}{{(2\pi)^4}} ~ (2\pi)^2 \delta(p^2-M_N^2)   \delta(\bar{p}^2-M_N^2)~\Omega(p,\bar{p}) = \nonumber \\
&&\qquad \qquad \left. \frac{1}{2\pi} \int d\xi \int \frac{d^2 \delta\pt}{(2\pi)^2}  \frac{1}{\sqrt{1-\xi^2}} \rho_{N}(t) \right|_{P^{+}=P^{+}_N\;,~\Pt=0}\;,
\end{eqnarray}
such that upon using the standard decomposition of the GPD between unpolarized hadron states \cite{Diehl:2003ny}
\begin{eqnarray}
q^{V}(x,\xi,t)=\sqrt{1-\xi^2} \left[ H(x,\xi,t) - \left(\frac{\xi}{x}\right)^2 \frac{x^2}{x^2-\xi^2} E(x,\xi,t)\right]\;, 
\end{eqnarray}
the current density can be written as 
\begin{eqnarray}
&& \begin{pmatrix} \lav j^{+}(y) \rav  \\ \lav  j^{i}(y) \rav  \\ \lav j^{-}(y) \rav  \end{pmatrix}
=\frac{1}{\pi}\int_{-\infty}^{\infty} dx  \int d\left(\frac{\xi}{x}\right) \int \frac{d^2\delta\pt}{(2\pi)^2}  \int \frac{d^2\Kt}{(2\pi)^2} \int \frac{d^2\dkt}{(2\pi)^2} \int d^2\bt \nonumber \\
&& \rho_{N}(t)~\left[ H(x,\xi,t) - \left(\frac{\xi}{x}\right)^2 \frac{x^2}{x^2-\xi^2} E(x,\xi,t)\right] D_{W}(\Kt,\bt)~\nn\\
&& \,\,\,\,\,\,\,\,\,\,\,\,\,\,\,\,\,\,\,\,\,\,\,\,\,\,\,\,\,\,\,\,\,\,\,\,\,\,\,\,\,\,\,\,\,\,\,\,\,\,\,\,\,
\begin{pmatrix} 
xP^{+} \\
\left( \Kt^{i} + \left( \frac{-\xi}{x} \right) \frac{\dkt^{i}}{2} \right) \frac{x^2}{x^2-\xi^2}\\
\frac{\Kt^2 - \frac{\dkt^2}{4} +m^2}{2xP^{+}}  \frac{x^2}{x^2-\xi^2}
\end{pmatrix} 
\nonumber \\
&&  e^{i\delta\pt\bt} e^{+2i  \left( \frac{-\xi}{x} \right) \left(x P^{+} y^{-} - \frac{\Kt^2+\frac{\dkt^2}{4}+m^2}{2xP^+}  \frac{x^2}{x^2-\xi^2} y^{+} \right)} e^{+i\dkt \left(\yt-\bt - \frac{\Kt}{xP^{+}}   \frac{x^2}{x^2-\xi^2} y^{+} \right)}\Big|_{P^{+}=P^{+}_N\;,~\Pt=0} \;. \nonumber \\
\label{crnt1}
\end{eqnarray}
Now the crucial observation is that the $t$ dependence of the integrand is in fact dominated by distribution of nucleus inside the nucleus, which severely restricts $|t| \lesssim 1/R_A^2$. However, on this scale variations of the $t$ dependence of the distributions of quarks inside the nucleon is suppressed by powers of $R_{p}/R_{A}$ and one may therefore approximate $t\simeq 0$ in the GPDs. Similar remarks can be made with regards to the $\xi$ dependence, which is kinematically related to the $t$ dependence via
\begin{eqnarray}
\left.t\right|_{\Pt=0}=-\frac{4 M_n^2 \xi^2}{1-\xi^2}-\frac{(\mathbf{\delta p})^2}{1-\xi^2}\;. \label{eq:59}
\end{eqnarray}
Since both terms in Eq.~\ref{eq:59} contribute with the same sign, one concludes that 
\begin{eqnarray}
\xi^2 < \frac{1}{1+4M_n^2/|t|} \lesssim \frac{1}{4M_n^2R_A^2}\;,
\end{eqnarray}
such that $|\xi| \lesssim 1/M_nR_A << 1$. Since on the other hand the GPDs typically vary on scales $|\xi| \sim x$, with the baryon charge dominated by the valence sector $x\sim1$, one may approximate $\xi \simeq 0$ up to corrections of order $R_{p}/R_{A}$, such that
\begin{eqnarray}
 && \begin{pmatrix} \lav j^{+}(y) \rav  \\ \lav  j^{i}(y) \rav  \\ \lav j^{-}(y) \rav  \end{pmatrix}
\simeq \frac{1}{\pi} \int_{-\infty}^{\infty} dx ~ H(x,0,0)  \int d\left(\frac{\xi}{x}\right) \int \frac{d^2\delta\pt}{(2\pi)^2} ~\rho_{N}(t) \nonumber \\
&&  \int \frac{d^2\Kt}{(2\pi)^2} \int \frac{d^2\dkt}{(2\pi)^2} \int d^2\bt ~D_{W}(\Kt,\bt)
\begin{pmatrix} 
xP^{+} \\
\Kt^{i} + \left( \frac{-\xi}{x} \right) \frac{\dkt^{i}}{2} \\
\frac{\Kt^2 - \frac{\dkt^2}{4} +m^2}{2xP^{+}}
\end{pmatrix} 
\nonumber \\
&& e^{i\delta\pt\bt} ~e^{+2i  \left( \frac{-\xi}{x} \right) \left(x P^{+} y^{-} - \frac{\Kt^2+\frac{\dkt^2}{4}+m^2}{2xP^+}  y^{+} \right)} e^{+i\dkt \left(\yt-\bt - \frac{\Kt}{xP^{+}}  y^{+} \right)}\Big|_{P^{+}=P^{+}_N\;,~\Pt=0} \;. \nonumber \\
\label{eq:jCurrentLargeNucl}
\end{eqnarray}
where by approximating $\xi\simeq0$ and $t\simeq0$ one recovers the collinear quark distribution $H(x,0,0)$, with negative values of $x$ denoting the corresponding anti-quark distribution.  Evaluating the integrals over $\delta \kt$ and $\bt$, by expressing $\delta\kt \to -i \nabla_{\yt}$ then yields
\begin{eqnarray}
\begin{pmatrix} \lav j^{+}(y) \rav  \\ \lav  j^{i}(y) \rav  \\ \lav j^{-}(y) \rav  \end{pmatrix}
&\simeq&  \frac{1}{\pi}  \int_{-\infty}^{\infty} dx~H(x,0,0)  \int d\left(\frac{\xi}{x}\right) \int \frac{d^2\delta\pt}{(2\pi)^2} ~\rho_{N}(t)  \nonumber \\
&&\int \frac{d^2\Kt}{(2\pi)^2}~D_{W}\left(\Kt,\yt- \frac{\Kt}{xP^{+}}  y^{+}\right)
\begin{pmatrix} 
xP^{+} \\
\Kt^{i} - i\left( \frac{-\xi}{x} \right) \frac{\nabla_{\yt}^{i}}{2} \\
\frac{\Kt^2 + \frac{\nabla_{\yt}^2}{4} +m^2}{2xP^{+}}
\end{pmatrix} 
\nonumber \\
&&e^{i\delta\pt \left(\yt- \frac{\Kt}{xP^{+}}  y^{+} \right)} e^{+2i  \left( \frac{-\xi}{x} \right) \left(x P^{+} y^{-} - \frac{\Kt^2-\frac{\nabla_{\yt}^2}{4}+m^2}{2xP^+}  y^{+} \right)}  \nonumber \\
\label{crnt2}
\end{eqnarray}
Neglecting further the gradients in impact parameter space ($\nabla_{\yt}$) which are suppressed by inverse powers of $Q_s R_A \gg 1$ relative to the leading result, the expression can be re-cast into the form
\begin{eqnarray}
\lav j^{\mu}(y) \rav = \int_{0}^{\infty} \frac{dk^{+}}{2k^{+}} \int \frac{d^2\kt}{(2\pi)^2} ~ 2k^{\mu}~ \Big[f_{q}(y,k) - f_{\bar{q}}(y,k)  \Big]_{k^2=m^2}
\end{eqnarray}
with a phase-space distribution
\begin{eqnarray}
f_{q/\bar{q}}(\yp,\ym,\yt,\kp,\kt)=f^{(0)}_{q/\bar{q}}\left(\ym - \frac{\km}{\kp} \yp,\yt- \frac{\kt}{\kp} \yp,\kp,\kt \right)\;.
\label{f}
\end{eqnarray}
given by
\begin{eqnarray}
f^{(0)}_{q}(\ym,\yt,\kp,\kt)&=& D_{W}(+\kt,\yt)~q(x) \int \frac{d\xi}{(2\pi)}  \int \frac{d^2\delta \pt}{(2\pi)^2}  \rho_{N}(t)~e^{-i\xi P^{+} \ym} e^{-i\delta \pt \cdot \yt}\;, \nonumber \\
f^{(0)}_{\bar{q}}(\ym,\yt,\kp,\kt)&=& D_{W}(-\kt,\yt)~\bar{q}(x) \int \frac{d\xi}{(2\pi)}  \int \frac{d^2\delta \pt}{(2\pi)^2}  \rho_{N}(t)~e^{-i\xi P^{+} \ym} e^{-i\delta \pt \cdot \yt}\;, \nonumber \\
\label{f0}
\end{eqnarray}
where we denote $x=\frac{k^{+}}{P^{+}}$ and used that $H(x,0,0)=q(x)$ for $x>0$ and $H(x,0,0)=-\bar{q}(x)$ for $x<0$. Since the distribution in Eq.~(\ref{f}), clearly satisfies the collisionless Boltzmann equation 
\begin{eqnarray}
k^{\mu}\partial_{\mu}^{y} f(y,k) =0\;,
\end{eqnarray}
one concludes that the final result can be interpreted in terms of classical particle picture, with the initial phase-space distribution determined by Eq.~(\ref{f0}). \\

\subsection{Space-time picture of baryon stopping in collisions of large nuclei}
We will now investigate the baryon compression and acceleration in more detail, and first consider the limit $\yp \to 0$, i.e. immediately after the collision. Considering for simplicity
\begin{eqnarray}
D_{W}(\Kt,\bt)=\frac{(2\pi)^2}{\pi Q_s^2(\bt)} \exp\left( -\frac{\Kt^2}{Q_s^2(\bt)} \right)\;,
\end{eqnarray}
one finds that
\begin{eqnarray}
 \lav j^{+}(\yp=0^{+},\ym,\yt)  \rav &=&\tilde{\rho}_{N}(\ym,\yt)~\int_{0}^{1} dx~\Big[q(x) -\bar{q}(x) \Big] \;, \\
 \lav j^{-}(\yp=0^{+},\ym,\yt)  \rav &=&\tilde{\rho}_{N}(\ym,\yt)~\int_{0}^{1} dx~\Big[q(x) -\bar{q}(x) \Big] ~\frac{Q_{s}^2(\yt)+m_q^2}{2(xP^{+})^2} \nonumber
\end{eqnarray}
where 
\begin{eqnarray}
 \tilde{\rho}_{N}(\ym,\yt) = P^{+} \int \frac{d\xi}{(2\pi)}  \int \frac{d^2\delta \pt}{(2\pi)^2}  \rho_{N}(t)~e^{-i\xi P^{+} \ym} e^{-i\delta \pt \cdot \yt}
\end{eqnarray}
is the density of nuclear matter inside the projectile. Note that due to Lorentz contract the distribution in the high-energy limit $P^{+} \to \infty$ becomes strongly peaked in $\xm$ as
\begin{eqnarray}
\lim_{P^{+} \to \infty}P^{+}  \int \frac{d\xi}{(2\pi)}~\rho_{N}\Big(t(\xi,\delta \pt)\Big) ~e^{-i\xi P^{+} \ym} = \rho_{N}\Big(t(\xi=0,\delta\pt)\Big)~\delta(\ym)\;.
\end{eqnarray}
Hence to estimate the baryon compression factor, it is important to extract the baryon density in the local rest-frame as discussed in Sec.~\ref{sqs}. Denoting
\begin{eqnarray}
\lav \frac{1}{x^2} \rav_{B} =\int_{0}^{1}  dx~\Big[q(x) -\bar{q}(x) \Big] \frac{1}{x^2}\;,
\end{eqnarray}
we find that
\begin{eqnarray}
 \lav n_{B}(\yp=0^{+},\ym,\yt)  \rav  &\simeq& \sqrt{\frac{Q_s^2(\yt)}{M_{N}^2}} \sqrt{\lav \frac{1}{x^2} \rav_{B}}  \lav n_{B}(\yp=0^{-},\ym,\yt)  \rav 
\end{eqnarray}
such that the typical compression factor is approximately given by
\begin{eqnarray}
\label{eq:CompressionEstimate}
\frac{ \lav n_{B} \rav_{\yp=0^{+}}}{ \lav n_{B} \rav_{\yp=0^{-}}} \approx 20 \left( \frac{Q_s^2}{25 {\rm GeV}} \right)^{1/2} \left( \frac{\lav \frac{1}{x^2} \rav_{B}}{0.06} \right)^{1/2}
\end{eqnarray}

\begin{figure}
\begin{center}
\includegraphics[width=0.95\linewidth]{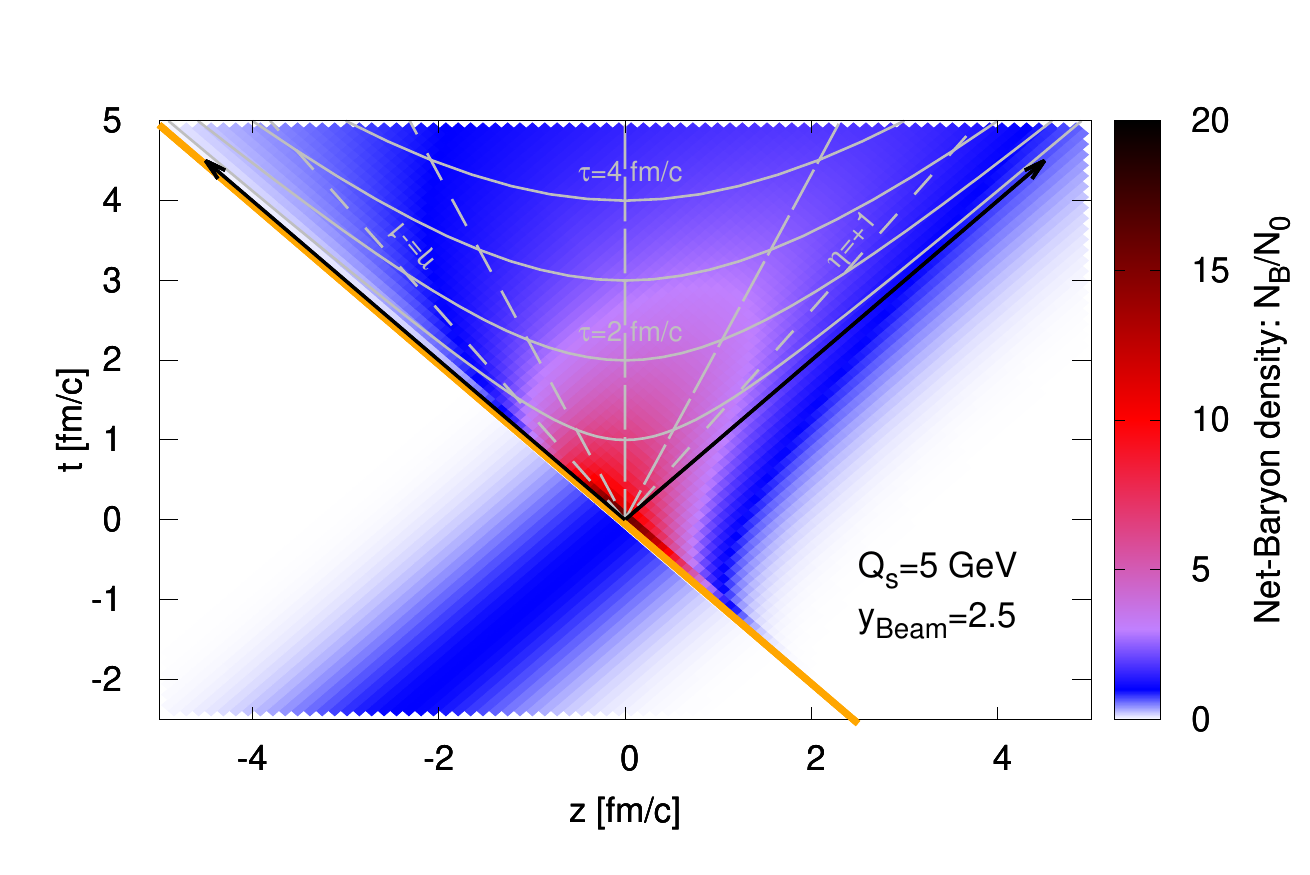}
\end{center}	
\caption{\label{fig:CurrentMap}Space time evolution of the baryon number density $N_B$ (in units of nuclear matter density $N_0=\frac{A}{(2\pi R_A^2)^{1/2}}$) before and after the collision of large nuclei. Numerical results are obtained for $R_A=6~{\rm fm}$ and $Q_s=5 {\rm GeV}$. In order to better illustrate the space-time dependence we have chosen a frame where the projectile nucleus has an initial momentum-space rapidity of $y_B=3$, such that the compressed and stopped baryonic matter is approximately at rest.}
\label{bstop}
\end{figure}

Beyond the limit $\yp\to0^{+}$ it becomes somewhat more involved to study the full (3+1)-dimensional space-time dynamics. We will therefore concentrate our attention to transverse averages of the  baryon currents
\begin{eqnarray}
J^{\mu}(\yp,\ym) = \int d^2\yt~j^{\mu}(\yp,\ym,\yt) \;. 
\end{eqnarray}
and for simplicity also neglect the impact parameter dependence in the target nucleus by setting $Q_s^2(\bt)=Q_s^2$. Starting from Eq.~(\ref{eq:jCurrentLargeNucl}), the corresponding expectation values of $J^{\mu}$ are then given by
\begin{eqnarray}
 \begin{pmatrix} \lav J^{+}(\yp,\ym)  \rav   \\ \lav J^{-}(\yp,\ym)  \rav  \end{pmatrix}
&\simeq& \int_{0}^{1} dx~\Big(q(x)-\bar{q}(x)\Big) \int d^2\Kt~\frac{e^{-\Kt^2/Q_s^2}}{\pi Q_s^2}
\begin{pmatrix} 
xP^{+} \\
\frac{\Kt^2 +m^2}{2xP^{+}}
\end{pmatrix} 
\nonumber \\
&&~ \frac{1}{\pi} \int d\left(\frac{\xi}{x}\right)~\rho_{N}\left(-\frac{4M_N^2\xi^2}{1-\xi^2}\right)~e^{+2i\left(\frac{-\xi}{x}\right) \left(xP^{+} y^{-} - \frac{\Kt^2+m^2}{2xP^+}  y^{+} \right)}   \;. \nonumber \\
\end{eqnarray}
Our results for the the space-time dependence of the baryon currents in the forward light-cone are compactly summarized in Fig.~\ref{fig:CurrentMap}, where we have evaluated this expression numerically, based on the following ansatz for the PDF and distribution of nuclear matter inside the projectile
\begin{eqnarray}
q(x)-\bar{q}(x)=\frac{\Gamma(a+b+2)}{\Gamma(a+1)\Gamma(b+1)} x^a (1-x)^b\;, \qquad \rho_{N}(t)=e^{-\frac{1}{2}R_A^2|t|}\;.
\end{eqnarray}
such that in coordinate space in the rest-frame of the nucleus $\rho_{N}(\vec{b})= \rho_0~e^{-\frac{1}{2}\vec{b}^2/R_A^2}$. Here $\rho_0=\frac{A}{(2\pi R_A^2)^{3/2}}$ denotes the nuclear matter density for a nucleus with radius $R_A$ and atomic number $A$, and if not stated otherwise, we use $a=b=3/2$ for the valence PDF parametrization.

Before the collision the nuclear matter inside the projectile is moving fast in the positive $z$ direction and the rest frame density $N_{B}=\sqrt{2 J^{+} J^{-}}$ is given by the typical nuclear matter density  $N_0=\frac{A}{(2\pi R_A^2)^{1/2}}$. Due to the interaction with the shock-wave the nuclear matter is decelerated and compressed, such that immediately after the interaction, i.e. $\yp=0^{+}$, the typical rest frame densities can reach several times nuclear matter density depending on the saturation scale $Q_s$ (c.f. Eq.~(\ref{eq:CompressionEstimate})). Specifically, for the values chosen in Fig.~\ref{fig:CurrentMap}, baryon densities of up to five times nuclear matter density persist for about $2~{\rm fm}/c$, within approximately one unit of rapidity concentrated around $\eta| \lesssim 0.5$ in the chosen frame. 

\begin{figure}
\begin{center}
\includegraphics[width=0.9\linewidth]{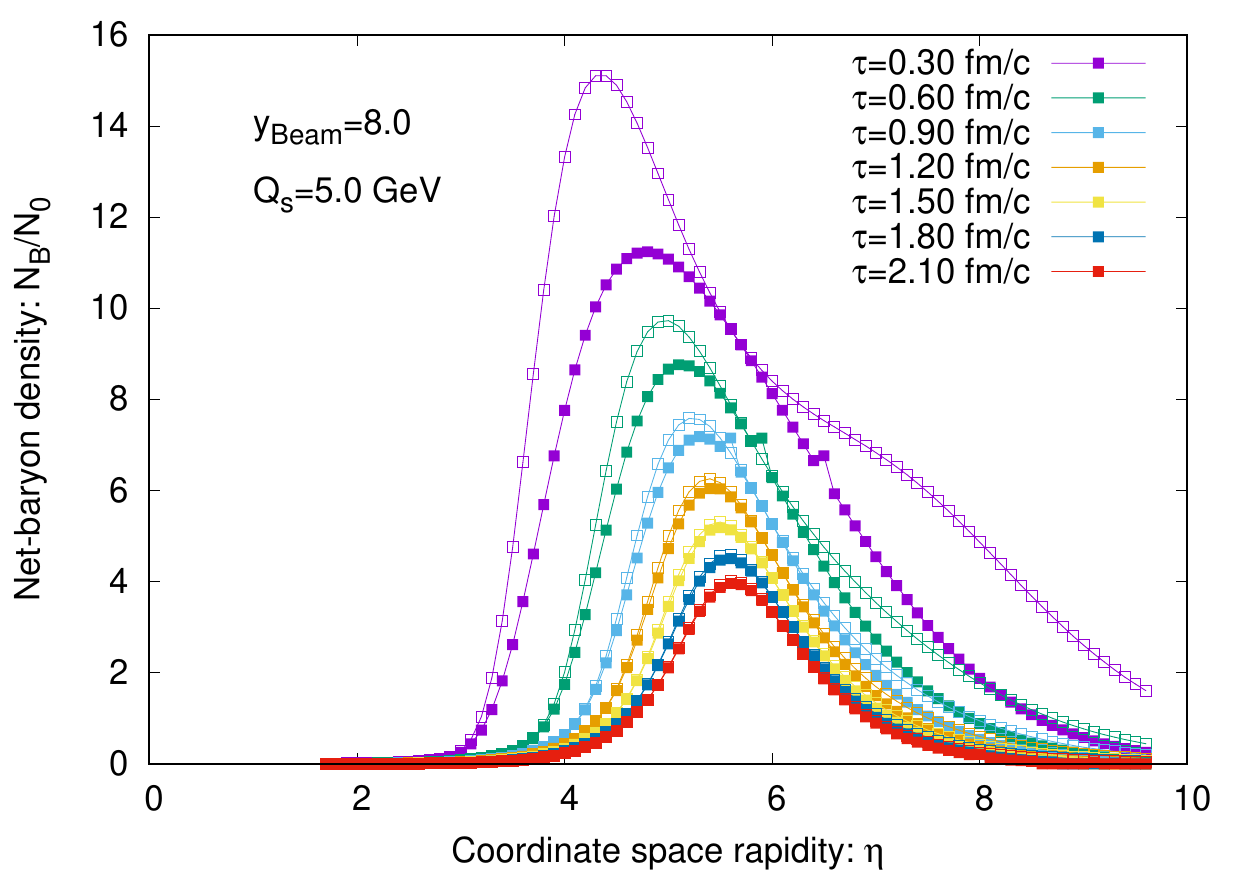}
\includegraphics[width=0.9\linewidth]{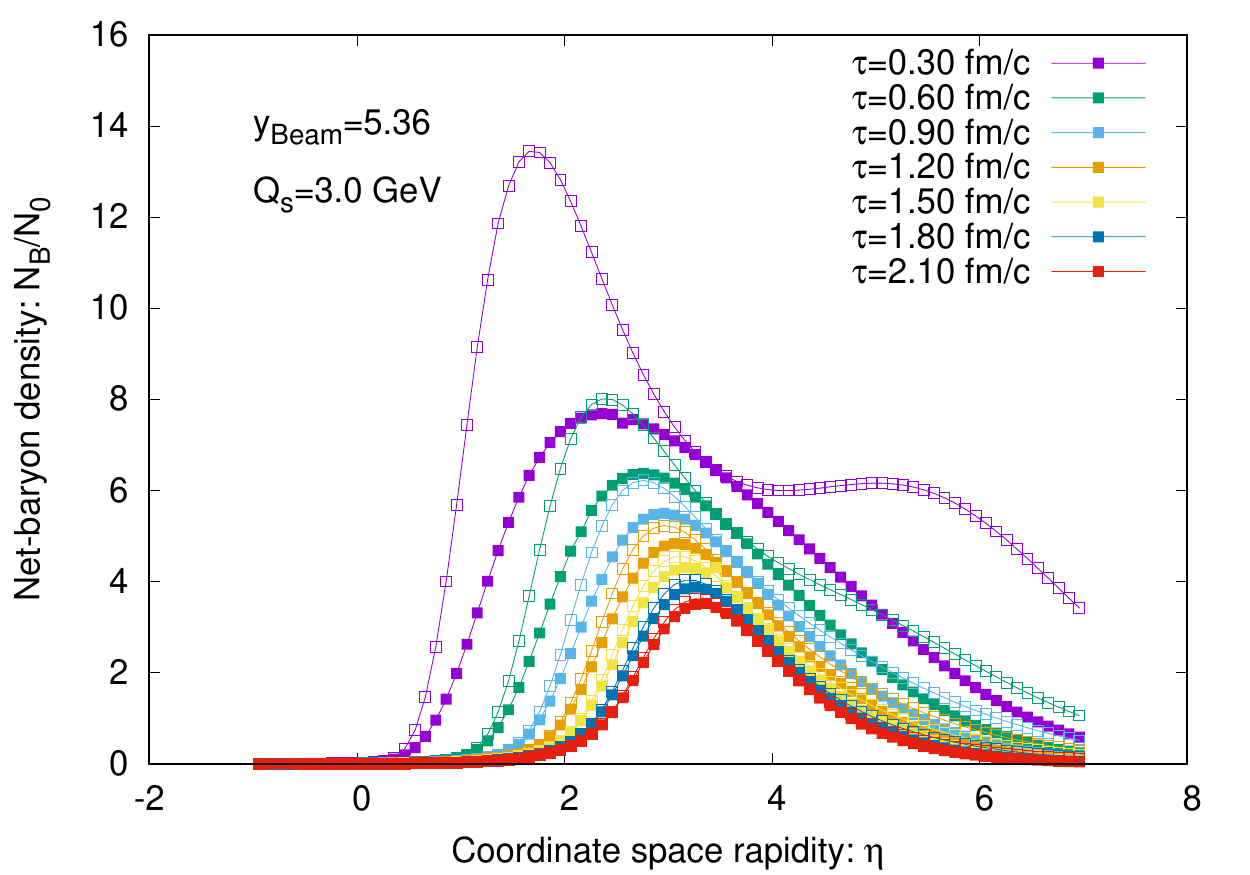}
\end{center}	

\caption{\label{fig:DensityVsRapidity}Net-baryon density profiles as a function of coordinate space rapidity $\eta$ at different times $\tau$ after the collision. Open symbol correspond to $J^{\tau}$, filled symbols show the results for the rest-frame density $N_{B}$.}
\end{figure}

We also note that we have chosen the origin of the coordinate system as the point where the center of mass of both nuclei coincide with each other. Since, unlike in the high-energy limit, the interaction region is extended in $\xm$, this also implies that at early times a significant fraction of the baryon charge is concentrated outside the forward light-cone. Hence to investigate the dynamics at early times, it is much more convenient to look at the dynamics in terms of $\xp,\xm$ or equivalently $t,z$ variables rather than the usual $\tau,\eta$ which are only defined in the forward light-cone. Nevertheless, one also observes from Fig.~\ref{fig:CurrentMap} that at later times $t\gtrsim 2 {\rm fm}/c$, most of baryon charge is concentrated in the forward light-cone, where one can then investigate profiles of the net-baryon density as a function of the coordinate space rapidity $\eta$. Our results are shown in Fig.~\ref{fig:DensityVsRapidity}, where we present the evolution of the rest-frame density $N_{B}$ in comparison with the density $J^{\tau}$ in a co-moving frame. Note that the two quantities behave quite differently at early times, due to the presence of additional currents $J^{\eta}$ in the longitudinal direction. Eventually, at late times $N_{B}$ and $J^{\tau}$ start to coincide with each other and approach the asymptotic form 
\begin{eqnarray}
&&\lim_{\tau \to \infty} \lav \tau J^{\tau}(\tau,\eta)  \rav \simeq   \int d^2\Kt~\frac{e^{-\Kt^2/Q_s^2}}{\pi Q_s^2}  \\
&& \qquad \left. \sqrt{\frac{\Kt^2+m_{q}^2}{M_{n}^2}} e^{\eta-y_{\rm Beam}} \Big[q(x)-\bar{q}(x)\Big]  \theta(1-x) \rho_{N}(t=0) 
\right|_{x=\sqrt{\frac{\Kt^2+m_{q}^2}{M_{n}^2}} e^{\eta-y_{\rm Beam}}} \nonumber
\end{eqnarray}
where the coordinate space profile of the projectile becomes irrelevant, and the net-baryon density exhibits the usual $1/\tau$ dependence characteristic of Bjorken expansion.

\newpage

\section{Summary and Conclusions}
\label{sec:conc}
In this paper we showed how to compute the distribution of valence particles after a collision with a high energy nucleus
when the nucleus is described by the theory of the Color Glass Condensate.  This demonstrates explicitly the compression of the valence
particles after the collision, along with simple estimates for the compression factors.  Quantum coherence plays a significant role
in the early time evolution of the scattered valence particles. However, in the limit of large colliding nuclei a semi-classical picture can be recovered, where so far the produced quarks are non interacting after the collision.  It remains to determine the gluon radiation induced by such a collision. As such, this work provide some initial steps and hints about how to treat the fragmentation region from first principles in QCD
at least for very high energy collisions.  Needless to say, there remains much work to be done before this goal is achieved.

\section{Acknowledgements}
We thank T.~Lappi, G.~Miller and R.~Venugopalan for discussions. This work is supported for L. McLerran  and Srimoyee Sen  under Department of Energy under grant number DOE grant No. DE-FG02-00ER41132 and for Soeren Schlichting in part under DE-FG02-97ER41014.
 
\appendix
\section{Baryon currents before the collision}\label{A}
Before the collision $(\yp<0)$ the baryon currents are given by
\begin{eqnarray}
\lav j^{\mu}(y)\rav &=& \int \frac{d^4p}{{(2\pi)^4}} \int \frac{d^4\bar{p}}{{(2\pi)^4}} ~ (2\pi)^2 \delta(p^2-M_N^2)   \delta(\bar{p}^2-M_N^2)~\Omega(p,\bar{p})  \nonumber \\
&&<\bar{p}| \PsiBar(y) \gamma^{\mu} \Psi(y) |p>\;,
\end{eqnarray}
which can be compactly expressed in terms of the form factors as
\begin{eqnarray}
\lav j^{\mu}(y)\rav = \int \frac{d^4p}{{(2\pi)^4}} \int \frac{d^4\bar{p}}{{(2\pi)^4}} ~ (2\pi)^2 \delta(p^2-M_N^2)   \delta(\bar{p}^2-M_N^2)~\Omega(p,\bar{p})  \nonumber \\
\bar{u}(\bar{p})\left[ F_{1}^{q}(t) \gamma^{+} +F_{2}^{q}(t) \frac{i \sigma^{+\alpha} (\bar{p}-p)_{\alpha}}{2M_{N}}  \right] u(p)~e^{-i  (\bar{p}-p) y}
\end{eqnarray}
Evaluating this expression in the limit of a large nucleus, i.e. for $\Omega(p,\bar{p})$ as in Eq.~(\ref{omega}) and approximating $t\simeq0$ and $\xi\simeq0$ one finds
\begin{eqnarray}
\lav j^{+}(y)\rav = \frac{1}{2\pi} \int d\xi \int \frac{d^2 \delta\pt}{(2\pi)^2} ~2P^{+}~\rho_{N}(t)F_{1}^{q}(t\simeq0) ~e^{-i  (\bar{p}-p) y}\;,
\end{eqnarray}
such that upon integration over the transverse coordinates $\yt$ the averaged current density is obtained as
\begin{eqnarray}
\lav J^{+}(\yp,\ym) \rav=P^{+}~\frac{1}{\pi} \int d\xi~\rho_{N}\left(\frac{-4M_N^2\xi^2}{1-\xi^2} \right) e^{-2i\xi \left(P^{+} \ym - \frac{M_N^2}{2P^{+}(1-\xi^2)} \yp\right)}\;.
\end{eqnarray}

\section{Early time limit $(y^{+}=0^{+})$ and relation to form factors}
Below we investigate more closely the limit $y^{+}=0^{+}$ of the baryon currents immediately after the collision, where several simplifications occur in Eq.~(\ref{curr1}) due to the absence of phase factors associated with the propagation in the forward light-cone
\begin{eqnarray}
&& \left. \lav j^{+}(y) \rav \right|_{y^{+}=0^{+}} =\int \frac{d^4p}{{(2\pi)^4}} \int \frac{d^4\bar{p}}{{(2\pi)^4}} ~ (2\pi)^2 \delta(p^2-M_N^2)   \delta(\bar{p}^2-M_N^2)~\Omega(p,\bar{p})  \nonumber \\
&&2P^{+}~\int_{-\infty}^{\infty} dx~q^{V}(x,\xi,t)~e^{+2 i  \xi P^{+} y^{-}} \int \frac{d^2\Kt}{(2\pi)^2} \int \frac{d^2\dkt}{(2\pi)^2} \int d^2\bt~D_{W}(\Kt,\bt)~\nn\\
&&\hspace{3.7in}e^{i(\bar{\pt}-\pt)\bt}  e^{-i\dkt (\bt-\yt)} \nonumber
\end{eqnarray}
Specifically, for the light-cone $+$ component of the current the relevant operator is trivial
\begin{eqnarray}
\int \frac{d^2\Kt}{(2\pi)^2}~D_{W}(\Kt,\bt) &=&\frac{1}{N_c} \tr[V_{\bt} V^{\dagger}_{\bt}]=1\;, \\
\end{eqnarray}
such that the interaction with the target initially has no effect 
\begin{eqnarray}
&&  \left. \lav j^{+}(y) \rav \right|_{y^{+}=0^{+}} = \int \frac{d^4p}{{(2\pi)^4}} \int \frac{d^4\bar{p}}{{(2\pi)^4}} ~ (2\pi)^2 \delta(p^2-M_N^2)   \delta(\bar{p}^2-M_N^2)~\Omega(p,\bar{p})  \nonumber \\
&&\int_{-\infty}^{\infty} dx  \int \frac{d^2\dkt}{(2\pi)^2} \int d^2\bt~2P^{+}~q^{V}(x,\xi,t)~e^{i(\bar{\pt}-\pt)\bt} e^{-2i  \xi P^{+} y^{-}} e^{-i\dkt (\bt-\yt)} \nonumber
\end{eqnarray}
which is identical to the expression obtained at $y^{+}=0^{-}$, i.e. immediately before the collision. Evaluating the expression explicitly by performing the integrations over $\dkt$ and $\bt$ one finds
\begin{eqnarray}
 \left. \lav j^{+}(y) \rav \right|_{y^{+}=0} &=& \int \frac{d^4p}{{(2\pi)^4}} \int \frac{d^4\bar{p}}{{(2\pi)^4}} ~ (2\pi)^2 \delta(p^2-M_N^2)   \delta(\bar{p}^2-M_N^2)~\Omega(p,\bar{p})  \nonumber \\
&&\bar{u}(\bar{p})\left[ F_{1}^{q}(t) \gamma^{+} +F_{2}^{q}(t) \frac{i \sigma^{+\alpha} (\bar{p}-p)_{\alpha}}{2M_{N}}  \right] u(p)~e^{-i  (\bar{p}-p) y} \nonumber \\
\end{eqnarray}
where we used the idenity \cite{Diehl:2003ny}
\begin{eqnarray}
2P^{+}~\int_{-\infty}^{\infty} dx~q^{V}(x,\xi,t)=\bar{u}(\bar{p})\left[ F_{1}^{q}(t) \gamma^{+} +F_{2}^{q}(t) \frac{i \sigma^{+\alpha} (\bar{p}-p)_{\alpha}}{2M_{N}}  \right] u(p)
\end{eqnarray}
relating the moment of the GPD to the Dirac and Pauli form factors $F_{1}^{q}(t)$ and $F_{2}^{q}(t)$ . 

  .

\end{document}